\documentclass[sigconf,10pt]{acmart}

\copyrightyear{2019} 
\acmYear{2019} 
\setcopyright{acmcopyright}
\acmConference[HotNets '19]{The 18th ACM Workshop on Hot Topics in Networks}{November 13--15, 2019}{Princeton, NJ, USA}
\acmBooktitle{The 18th ACM Workshop on Hot Topics in Networks (HotNets '19), November 13--15, 2019, Princeton, NJ, USA}
\acmPrice{15.00}
\acmDOI{10.1145/3365609.3365852}
\acmISBN{978-1-4503-7020-2/19/11}

\usepackage{times}  
\usepackage{hyperref}

\hypersetup{pdfstartview=FitH,pdfpagelayout=SinglePage}
\usepackage{booktabs}  
\usepackage{pbox}
\usepackage{subfigure}
\usepackage{epsfig,endnotes}
\usepackage{epstopdf}
\usepackage{xcolor}
\usepackage{graphicx}
\usepackage{xspace}
\usepackage{enumitem}
\usepackage{url}
\usepackage{flushend}  
\usepackage{arydshln}

\newcommand{\eg}{{e.g.,}\xspace}
\newcommand{\ie}{{\it i.e.,}\xspace}
\newcommand{\folder}{.}

\newcommand\co[1]{}

\newcommand{\pl}{{PlanetLab}\xspace}
\newcommand{\tool}{{BatteryLab}\xspace}
\newcommand{\wifi}{{WiFi}\xspace}

\smallskip
\smallskip

\begin{document}

\title[BatteryLab]{BatteryLab, A Distributed Power Monitoring Platform For Mobile Devices}
\subtitle{\url{https://batterylab.dev}}
\author{Matteo~Varvello$\dag$, Kleomenis~Katevas$\diamond$, Mihai~Plesa$\dag$, Hamed~Haddadi$\dag\diamond$, Benjamin~Livshits$\dag\diamond$}
\affiliation{%
  \institution{$\dag$~Brave~Software, $\diamond$~Imperial~College~London}
}

\begin{abstract}
Recent advances in cloud computing have simplified the way that both software  development and testing are performed. Unfortunately, this is not true for battery testing for which state of the art test-beds simply consist of one phone attached to a power meter. These test-beds have limited resources, access, and are overall hard to maintain; for these reasons, they often sit idle with no experiment to run. In this paper, we propose to \emph{share} existing battery testing setups and build \tool, a distributed platform for battery measurements. Our vision is to transform independent battery testing setups into \emph{vantage points} of a planetary-scale measurement platform offering heterogeneous devices and testing conditions. In the paper, we design and deploy a combination of hardware and software solutions to enable \tool's vision. We then preliminarily evaluate \tool's accuracy of battery reporting, along with some system benchmarking. We also demonstrate how \tool can be used by researchers to investigate a simple research question. 
\end{abstract}

\renewcommand{\shortauthors}{Varvello~et~al.}
\renewcommand{\authors}{Matteo~Varvello, Kleomenis~Katevas, Mihai~Plesa, Hamed~Haddadi, Benjamin~Livshits}

\maketitle

\section{Introduction}
\label{sec:intro}
The mobile device ecosystem is large, ever growing, and very much ``location-based'', \ie different devices and operating systems (Android and iOS) are popular at different locations. Advances in cloud computing have simplified the way that mobile apps are tested, today. Device \emph{farms}~\cite{awsfarm, appcenter} let developers test apps across a plethora of mobile devices, in real time.  Device diversity for testing is paramount since hardware and software differences might impact how an app is displayed or performs. 

To the best of our knowledge, no existing device farm offers \emph{hardware-based} battery measurements, where the power drawn by a device is measured by directly connecting its battery to an external power meter. Instead, few startups~\cite{greenspector,mobileenerlytics} offer \emph{software-based} battery measurements where device resource monitoring (screen, CPU, network, etc.) are used to infer the power consumed by few devices for which a calibration was possible~\cite{chenSIGMETRICS15}. This suggests a demand for battery measurements, but a prohibitive cost for deploying hardware-based solutions. 

In the research community, hardware-based battery measurements are instead quite popular~\cite{buiMOBICOM15, caoPOMAC17, ravenMOBICOM17,thiagarajanWWW12}. The common research approach consists of buying the required hardware (often an  Android device and a Monsoon power monitor~\cite{monsoon}), set it up on a desk, and then use it sporadically. This is because such battery testbeds are intrinsically \emph{local}, \ie they require a researcher or an app tester to have physical access to the device and the power meter. 

In this paper, we challenge the assumption that a battery testbed needs to be local and propose \emph{\tool}~\cite{batterylab}, a distributed platform for battery measurements. Similarly to \pl~\cite{planetlab}, our vision is a platform where members contribute hardware resources (\eg some phones and power monitor) in exchange of access to the hardware resources offered by other platform members. As new members join over time and from different locations, \tool will naturally grow richer of new and old devices, as well as of devices only available at some specific locations.

\tool's architecture consists of an \emph{access server} --- which enables an end-to-end test pipeline while supporting multiple users and concurrent timed sessions --- and several \emph{vantage points}, \ie the local testbeds described above. Vantage points are enhanced with a lightweight \emph{controller} --- hosted on a Raspberry Pi~\cite{rasbpi} --- which runs \tool's software suite to enable remote testing, \eg SSH channel with the access server and \emph{device mirroring}~\cite{scrcpy} which provides full remote control of test devices, via the browser.

We first evaluate \tool with respect to the \emph{accuracy} of its battery readings. This analysis shows that the required extra \tool's hardware has negligible impact on the power meter reporting. It also shows a non-negligible cost associated with device mirroring, suggesting that it should only be used when devising a test. Such \emph{headless} mode is not always possible, \eg if usability testing is the goal. In this case, the extra battery consumption associated with mirroring should be accounted for. 

Finally, we demonstrate \tool usage investigating a simple research question: \emph{which of today's Android browser is the most energy efficient?} To answer this question, we automated the testing of four popular browsers (Chrome, Firefox, Edge, and Brave) via \tool. Our results show that Brave offers minimal battery consumption, while Firefox tends to consume the most. We further augment this result across multiple locations (South Africa, China, Japan, Brazil, and California) emulated via VPN tunneling. 

\section{Related Work}
\label{sec:related}
This work was mainly motivated by the frustration of not finding a tool offering easy access to battery measurements. Several existing tools could leverage some of \tool's ideas to match our capabilities in a paid/centralized fashion. For example, device farms such as AWS Device Farm~\cite{awsfarm} and Microsoft AppCenter~\cite{appcenter} could extend their offer using our hardware and software components. The same is true for startups like GreenSpector~\cite{greenspector} and Mobile Enerlytics~\cite{mobileenerlytics}, which offer software-based battery testing on few devices. 

To the best of our knowledge, MONROE~\cite{alay2017experience} is the only measurement platform sharing some similarities with \tool. This is a platform for experimentation in operational mobile networks in Europe. MONROE currently has presence in 4 countries with 150 \emph{nodes}, which are ad-hoc hardware configurations~\cite{monroe_node} designed for cellular measurements. \tool is an orthogonal measurement platform to MONROE since it targets real devices (Android and iOS) and fine-grained battery measurements. The latter requires specific instrumentation (bulky power meters) that cannot be easily added to MONROE nodes, especially the mobile ones. In the near future, we will explore solutions like BattOr~\cite{schulman2011phone} to potentially enhance \tool with mobility support. 

Last but not least, \tool offers full access to test devices via mirroring. This feature was inspired by~\cite{almeida2018chimp}, where the authors build a platform to allow an Android emulator to be accessed via the browser, with the goal to ``crowdsource'' human inputs for mobile apps. We leverage the same concept to allow remote access to \tool, but also further extend it to actual devices and not only emulators.

 \begin{figure*}[htb]
     \centering
     \hspace{-0.5in}    
     \subfigure[Distributed architecture.]{\psfig{figure=\folder/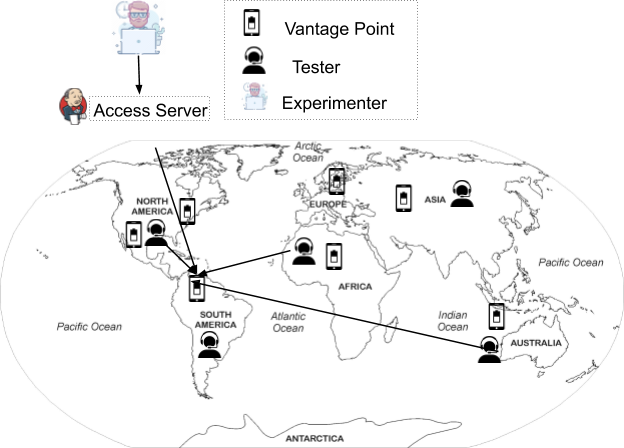, width=3in}\label{fig:arch-a}}
     \subfigure[Vantage point design.]{\psfig{figure=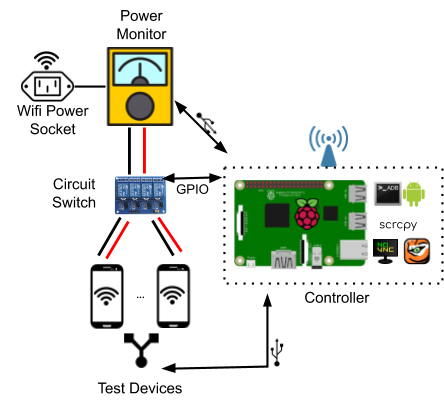, width=2.2in}\label{fig:arch-b}}
     \subfigure[GUI.]{\psfig{figure=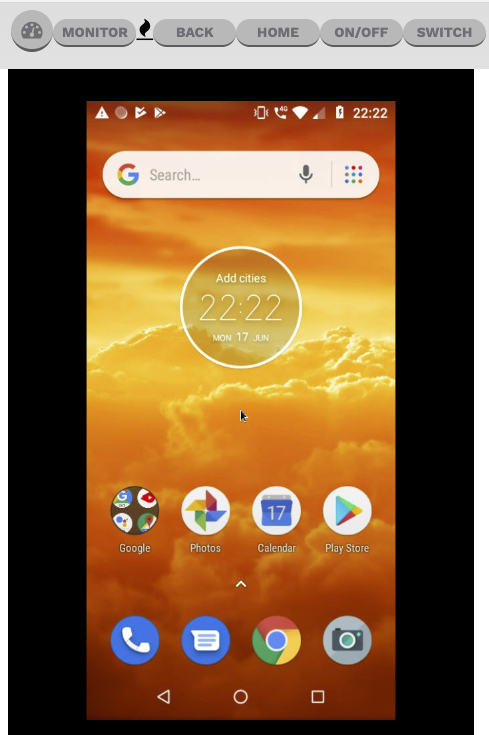, width=1.2in}\label{fig:gui}}
     \caption{\tool's infrastructure.}
     \label{fig:arch}
 \end{figure*}

\section{BatteryLab}
\label{sec:system}
This section details the design and implementation of \tool, a distributed measurement platform for device battery monitoring (see Figure~\ref{fig:arch-a}). We currently focus on mobile devices only, but our architecture is flexible and we thus plan to extend to more devices, \eg laptops and IoT devices.

One or multiple test devices (a phone/tablet connected to a power monitor) are hosted at some university or research organization around the world (\emph{vantage points}). \tool members (\emph{experimenters}) gain access to test devices via a centralized \emph{access server}, where they can request time slots to deploy automated scripts and/or ask remote control of the device. Once granted, remote control of the device can be shared with \emph{testers}, whose task is to manually interact with a device, \eg search for several items on a shopping application. Testers are either volunteers, recruited via email or social media, or paid, recruited via crowdsourcing websites like Mechanical Turk~\cite{mturk} and Figure Eight~\cite{figureeight}.

\subsection{Access Server}
\label{sec:sys:server}
The main role of the access server is to manage the vantage points and schedule experiments on them based on experimenter requests. We built the access server atop of the Jenkins~\cite{jenkins} continuous integration system which is free, open source, portable (as it is written in Java) and backed by an active and large community. Jenkins enables end-to-end test pipelines while supporting multiple users and concurrent timed sessions.

\tool's access server runs in the cloud (Amazon Web Services) which enables further scaling and cost optimization. Vantage points have to be added explicitly and pre-approved in multiple ways (IP lockdown, security groups). Experimenters need to authenticate and be authorized to access the web console of the access server. For increased security, this is only available over HTTPS.

The access server communicates with the vantage points via SSH. New \tool members grant SSH access from the server to the vantage point's controller via public key and IP white-listing (see Section~\ref{sec:sys:join}). Experimenters can access vantage points via the access server, where they can create \emph{jobs} to be deployed in their favorite programming language. Only the experimenters that have been granted access to the platform can create, edit or run jobs and every pipeline change has to be approved by an administrator. This is done via a role-based authorization matrix.

\tool's Python API (see Table~\ref{tab:api}) is available to provide user-friendly  device selection, interaction with the power meter, etc.  The access server will then dispatch queued jobs based on experimenter constraints, \eg target device, connectivity, or network location, and \tool constraints, \eg one job at the time per device.  By default, the access server collects logs from the power meter which are made available for several days within the job's workspace. Android logs like \texttt{logcat} and \texttt{dumpsys} can be requested via the \texttt{execute\_adb}  API, if available.

We have developed several jobs which manage the vantage points. These jobs span from updating \tool wildcard certificates (see Section~\ref{sec:sys:join}), to ensure the power meter is not active when not needed (for safety reasons), or to factory reset a device. These jobs were motivated by our needs while building the system, and we expect more to come over time and as the system grows.

\subsection{Vantage Point}
\label{sec:sys:remote}
Figure~\ref{fig:arch-b} shows a graphical overview of a \tool's vantage point with its main components: controller, power monitor, test devices, circuit switch, and power socket. 

\vspace{0.1in}
\noindent\textbf{Controller} -- This is a Linux-based machine responsible for managing the vantage point. The machine should be equipped with both Ethernet and \wifi connectivity, a USB controller with a series of available USB ports, as well as with an external General-Purpose Input/Output (GPIO) interface. We use the popular Raspberry Pi 3B+~\cite{rasbpi} running the latest version of Raspbian Stretch (April 2019) that meets these requirements with an affordable price.

The controller's primary role is to manage connectivity with test devices. Each device is connected to the controller's USB port, \wifi access point (configured in NAT or Bridge mode), and Bluetooth. USB connectivity is used to power each testing device when not connected to the power monitor and to instrument it via the Android Debugging Bridge~\cite{adb} (ADB), if available. \wifi connectivity is used to allow automation without the extra USB current, which interferes with the power monitoring procedure. (De)activation of USB ports is realized using \texttt{uhubctl}~\cite{uhubctl}. Bluetooth connectivity is used for automation across OSes (Android and iOS) and connectivity (\wifi and cellular). Section~\ref{sec:sys:automation} will discuss several automation techniques supported by \tool.

The second role of the controller is to provide \emph{device mirroring}, \ie easy remote access to the device under test. We use VNC (\texttt{tigervnc}~\cite{tigervnc}) to enable remote access to the controller. We further use \texttt{noVNC}~\cite{novnc}, an HTML VNC library and application, to provide easy access to a VNC session via a browser without no further software required at an experimenter or tester. We then \emph{mirror} the test device within the noVNC/VNC session and limit access to only this visual element. In Android, this is achieved using \texttt{scrcpy}~\cite{scrcpy}, a screen mirroring utility which runs atop of ADB. No equivalent software exists for iOS, but a similar  functionality can be achieved combining AirPlay Screen Mirroring~\cite{AirPlay} with (virtual) keyboard keys (see Section~\ref{sec:sys:automation}).

Figure~\ref{fig:gui} shows a snapshot of the graphical user interface (GUI) we have built around the default \texttt{noVNC} client. The GUI consists of an \emph{interactive area} and a \emph{toolbar}. The interactive area (bottom of the figure) is the area where a device screen is mirrored. As a user (experimenter or tester) hovers his/her mouse within this area, (s)he gains access to the device currently being mirrored, and each action is executed on the physical device. The GUI connects to the controller's backend using AJAX calls to some internal restful APIs. The toolbar occupies the top part of the GUI, and implements a convenient subset of \tool's API (see Table~\ref{tab:api}). Even though the toolbar was initially thought as a visual helper for an \emph{experimenter}, it is also useful for less experienced \emph{test participants}. For this reason, \tool allows an experimenter to control the presence or not of the toolbar on the webpage to be shared with a test participant.

\vspace{0.1in}
\noindent\textbf{Power Monitor} -- This is a power metering hardware capable of measuring the current consumed by a test device in high sampling rate. \tool currently supports the Monsoon HV~\cite{monsoon}, a power monitor with a voltage range of 0.8V to 13.5V and up to 6A continuous current sampled at $5KHz$. The Monsoon HV is controlled using its Python API~\cite{pymonsoon}. Other power monitors can be supported, granted that they offer APIs to be integrated with \tool's software suite.

\vspace{0.1in}
\noindent\textbf{Test Device(s)} -- It is a mobile device (phone or tablet) that can be connected to a power monitor. While we recommend phones with removable batteries, more complex setups requiring to (partially) tear off a device to reach the battery are possible. Note that, on Android, device mirroring is only supported on devices running  API 21 (Android $\geq$ 5.0). 

\vspace{0.1in}
\noindent\textbf{Circuit Switch} -- This is a relay-based circuit with multiple channels that lies between the test devices and the power monitor. The circuit switch is connected to the controller's GPIO interface and all relays can be controlled via software from the controller. Each relay uses the device's voltage (+) terminal as an input, and programmatically switches between the battery's voltage terminal and the power monitor's \texttt{Vout} connector. Ground (-) connector is permanently connected to all devices' Ground terminals.

This circuit switch has two main tasks. First, it allows to switch between a direct connection between the phone and its battery, and the ``battery bypass''---which implies disconnecting the battery and connecting to the power monitor. This is required to allow the power monitor to measure the current consumed during an experiment. Second, it allows \tool to concurrently support multiple test device without having to manually move cables around.

\vspace{0.1in}
\noindent\textbf{\wifi Power Socket} --  This is used to allow the controller to turn the Monsoon on and off, when needed. It connects to the controller via \wifi and it is controlled with some simple API. The current \tool software suite only supports Meross power sockets by integrating the following APIs~\cite{meross}. In the near future we will replace this power socket by extending the capabilities of the circuit switch.

\begin{table}[t]
\small
\centering
\begin{tabular}{rccc}
\hline
    {\bf API}     & {\bf Description}          & {\bf Parameters}  \\
    \hline
    {\bf list\_devices}       & \begin{tabular}{@{}c@{}}List ADB ids \\ of test devices\end{tabular}  & -             \\
    \hdashline[0.5pt/5pt]
    {\bf device\_mirroring}   & \begin{tabular}{@{}c@{}}Activate device \\ mirroring\end{tabular}    & device\_id    \\
    \hdashline[0.5pt/5pt]
    {\bf power\_monitor}      & \begin{tabular}{@{}c@{}}Toggle Monsoon\\ power state\end{tabular}   & -             \\
    \hdashline[0.5pt/5pt]
    {\bf set\_voltage}        & \begin{tabular}{@{}c@{}}Set target\\ voltage \end{tabular}          & voltage\_val  \\
    \hdashline[0.5pt/5pt]
    {\bf start\_monitor}      & \begin{tabular}{@{}c@{}}Start battery\\ measurement \end{tabular}    & device\_id, duration\\
    \hdashline[0.5pt/5pt]
    {\bf stop\_monitor}       & \begin{tabular}{@{}c@{}}Stop battery\\ measurement \end{tabular}     & -             \\
    \hdashline[0.5pt/5pt]
    {\bf batt\_switch}        & \begin{tabular}{@{}c@{}}(De)activate\\ battery \end{tabular}         & device\_id    \\
    \hdashline[0.5pt/5pt]
    {\bf execute\_adb}        & \begin{tabular}{@{}c@{}}Execute ADB\\ command \end{tabular}         & device\_id, command \\
    \hdashline[0.5pt/5pt]
    
\end{tabular}
\caption{\tool's API.}
\label{tab:api}
\end{table}

\subsection{Automation}
\label{sec:sys:automation}

\tool supports three mechanisms for test automation, each with its own set of advantages and limitations.

\vspace{0.1in}
\noindent \textbf{Android Debugging Protocol} (Android) --  ADB~\cite{adb} is a powerful tool/protocol to control an Android device. Commands can be sent over USB, \wifi, or Bluetooth. While USB guarantees highest reliability, it  interferes with the power monitor due to the power sent to activate the USB micro-controller at the device. This is solved by sending commands over \wifi or Bluetooth. However, using \wifi implies not being able to run experiments leveraging the mobile network, and ADB-over-Bluetooth requires a rooted device. Based on an experimenter needs, \tool can dynamically switch between the above automation solutions.

\vspace{0.1in}
\noindent \textbf{UI Testing} (Android and iOS) -- This solution uses UI testing frameworks (\eg~Android's user interface tests~\cite{android_tests} or Apple's XCTest framework~\cite{xctest}), to produce a separate version of the testing app, configured with automated actions. The advantage of this solution, compared with ADB, is that it does not require a communication channel with the Raspberry Pi. The main drawback is that it restricts the set of applications that can be tested since access to an app source code is required.

\begin{figure}[t]
    \centering
    \psfig{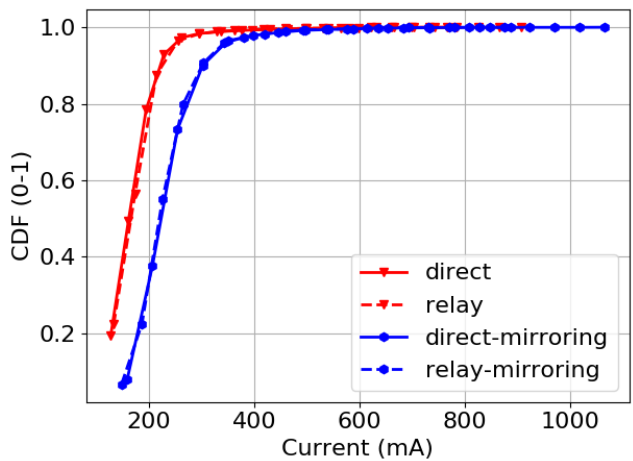}
    \caption{CDF of current drawn \newline(direct, relay, direct-mirroring, relay-mirroring).}
    \label{fig:relay}
 \end{figure}

\vspace{0.1in}
\noindent \textbf{Bluetooth Keyboard} (Android and iOS) -- This approach automates a test device via (virtual) keyboard keys (\eg locate an app, launch it, and interact with it). The controller emulates a typical keyboard service to which test devices connect via Bluetooth. This approach is generic and thus works for both Android and iOS devices, with no rooting needed. Since it relies on Bluetooth, it also enables experiments on the cellular network. The limitations are twofold. First, Android device mirroring is not supported as it requires ADB. This is not an issue for automated tests which can and should be run in \emph{headless} mode to minimize noise on the battery reporting (see Figure~\ref{fig:relay}). It follows that this limitation only applies to usability testing (with real users) on a mobile network.

The second limitation is that the level of automation depends both on the OS and app support for keyboard commands. In Android, it can be challenging to match ADB's API with this approach. It should be noted though that, when available, ADB can still be used ``outside'' of a battery measurement. That is, operations needed before and after the actual battery measurement (\eg cleaning an app cache) can still be realized using ADB over USB. When the actual test starts, \eg launch an app and perform some simple interactions, we can then switch to Bluetooth keyboard automation.

\subsection{How to Join?}
\label{sec:sys:join}
Institutions interested in joining \tool can do so by following our tutorial~\cite{blabtutorial}. In short, we recommend the hardware to use and its setup.  It is important for the controller to be publicly reachable at the following configurable ports: 2222 (SSH, access server only), 8080 (GUI's backend), 6081 (noVNC). Members will provide a human readable identifier for the vantage point which will be added to \tool's DNS (\eg \texttt{node1.batterylab.dev}) provided by Amazon Route53~\cite{route53}. Our wildcard \texttt{letsencrypt}~\cite{letsencrypt} certificate will be provided at this point. Renewal of this certificate is managed by the access server which also automatically deploys it at each vantage point, when needed.

The next step consists of flashing the controller (Raspberry Pi) with \tool's image. This will setup the most recent Raspbian version, along with \tool's required code and its configuration. Few manual steps are required to verify connectivity, grant pubkey access to the access server, and connect at least one Android device. At this point, the controller should be visible at the access server, and the device accessible at \texttt{https://node1.batterylab.dev}. 

\section{Preliminary Evaluation}
\label{sec:eval}
This section preliminarily evaluates \tool using its first vantage point deployed at Imperial College London, UK. 
This consists of a Monsoon power meter, a Samsung J7 Duo (Android 8.0), a Raspberry Pi 3B+, and a Meross power socket. We first evaluate \tool's \emph{accuracy} in battery measurements reporting. Next, we demonstrate its usage investigating a simple research question. We further use this demonstration to benchmark \tool's system performance. Finally, we experiment with the impact of multiple device \emph{locations} emulated via a VPN.

\subsection{Accuracy}
\label{sec:eval:accuracy}
Compared to a classic \emph{local} setup for battery measurements, \tool introduces some hardware (circuit relay) and software (device mirroring) components that can impact the \emph{accuracy} of the measurements collected. We devised a simple experiment where we compare three scenarios. First, a \emph{direct} scenario consisting of just the Monsoon power meter, the testing device, and the Raspberry Pi to instrument the power meter. For this setup, we strictly followed Monsoon indications~\cite{monsoon} in terms of tape, cable type and length, and connectors to be used. Next, we introduce two additional scenarios: a \emph{relay} scenario where the relay circuit is used to enable \tool's programmable switching between multiple devices as well as between battery bypass and regular battery operation (see Section~\ref{sec:sys:remote}). Finally, a \emph{mirroring} scenario where the device screen is mirrored to an open noVNC session. While the relay is always ``required'' for \tool to properly function, device mirroring is only required for usability testing. 

Figure~\ref{fig:relay} shows the Cumulative Distribution Function (CDF) of the current consumed in each of the above scenarios during a 5 minutes test. For completeness, we also consider a \emph{direct-mirroring} scenario where the device is directly connected to Monsoon and screencasting is active. During the test, we play an mp4 video pre-loaded on the device sdcard. The rationale is to force the device mirroring mechanism to constantly update as new frames are originated. The figure shows negligible difference between the ``direct'' and ``relay'' scenario, regardless of the device mirroring being active or not. A larger gap (median current grows from 160 to 220mA) appears with device mirroring. This is because of the background process responsible of screencasting to the controller which causes additional CPU usage on the device (Figure~\ref{fig:res-a}). 

 \begin{figure}[t]
    \centering
    \psfig{figure=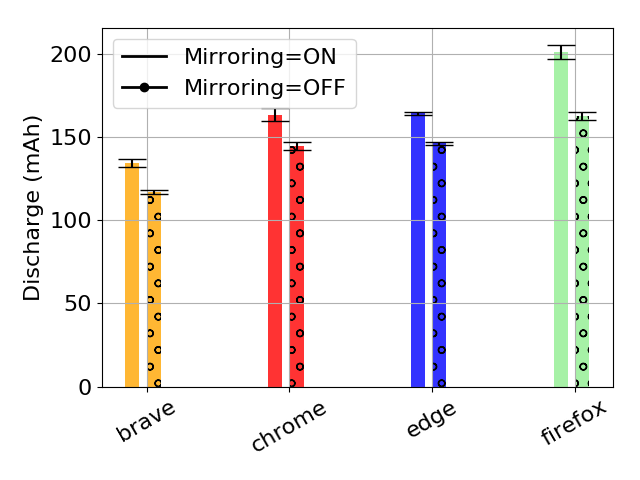, width=\columnwidth}
    \caption{Per browser energy consumption\newline(Brave, Chrome, Edge, Firefox).}
    \label{fig:res-c}
 \end{figure}

\subsection{Demonstration}
\label{sec:eval:demo}
We demonstrate \tool's usage assuming an experimenter asks the following question: \emph{which of today's Android web-browsers is the most energy efficient?} The experimenter writes an automation script which instruments a browser to load a webpage and interact with it. Scripts are deployed via \tool's Jenkins interface, and phone access is granted via device mirroring in the experimenter's browser. When satisfied with the automation, the experimenter can launch a real test with active battery monitoring. The experiment is added to Jenkin's queue and will run when the right conditions are met, \ie no other test is running (required) and low CPU utilization (optional). When an experiment completes, logs can be  retrieved via the Jenkins interface.

We build browser automation using bash and \tool's ADB over \wifi automation procedure. We automate a few popular Android browsers: Chrome, Firefox, Edge, and Brave. Our experiments are \wifi only since the device under test is not rooted. 
Each browser is instrumented to sequentially load $10$ popular news websites. After a URL is entered, the automation script waits 6 seconds -- emulating a typical page load time (PLT) for these websites under our (fast) network conditions -- and then interact with the page by executing multiple ``scroll up'' and ``scroll down'' operations. Before the beginning of a workload, the browser state is cleaned and the required setup is done, \eg Chrome requires at first launch to accept some conditions, sign-in into an account or not, etc. We iterate through each browser sequentially, and re-test each browser 5 times. We repeat the full experiment with both active and inactive device mirroring.

  \begin{figure}[t]
    \centering
    \psfig{figure=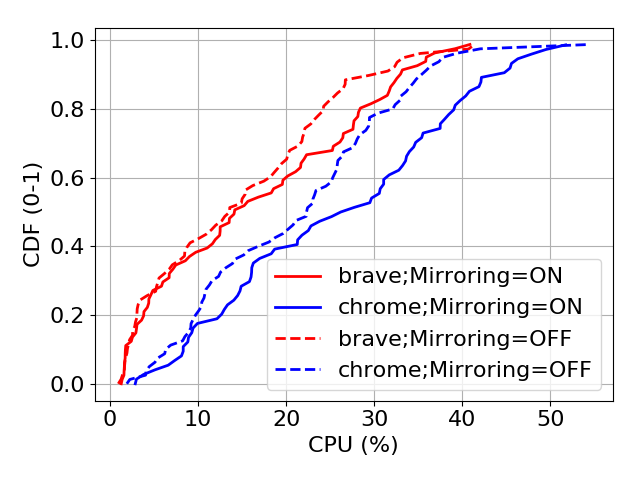, width=\columnwidth}
    \caption{CDF of CPU consumption\newline(Brave and Chrome).}
    \label{fig:res-a}
 \end{figure}

\vspace{0.1in}
\noindent\textbf{Browser Performance} 
Figure~\ref{fig:res-c} shows the average battery discharge (standard deviation as errorbars) measured for each browser, considering both active and inactive device mirroring. The figure shows that, regardless of device mirroring, the overall result does not change, \ie with Brave offering minimal battery consumption and Firefox consuming the most. This is because device mirroring offers a constant extra cost ($\sim$20mAh) regardless of the browser being tested. This result is in line with the constant gap observed between active and inactive device mirroring in Figure~\ref{fig:relay}. 

This additional battery consumption caused by device mirroring is due to an increase of the CPU load on the device under test. Figure~\ref{fig:res-a} shows the CDF of the CPU utilization for Chrome and Brave with active and inactive device mirroring, respectively. A similar trend is observed for the other browsers, which have been omitted to increase the plot visibility. The figure shows two results. First, Brave's lower battery consumption comes from an overall lower CPU pressure, \eg a median CPU utilization of 12\% versus 20\% in Chrome. Second, device mirroring causes, for both browsers, a 5\% CPU increase. This is more noticeable at higher CPU values which is when the browser's automation is active. This happens because of the increasing load on the encoder when the screen content changes quickly versus, for example, the fixed phone's home screen. 

  \begin{figure}[t]
    \centering
    \psfig{figure=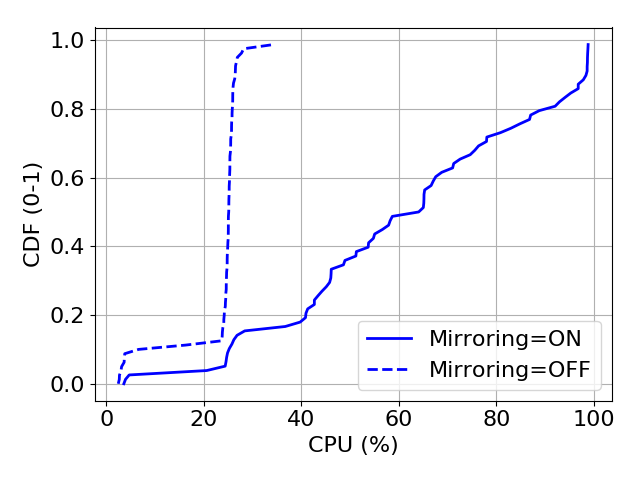, width=\columnwidth}
    \caption{CDF of CPU consumption at the controller (Raspberry Pi 3B+).}
    \label{fig:res-b}
 \end{figure}

\vspace{0.1in}
\noindent\textbf{System Performance}  
Overall, higher CPU utilization is the main extra cost caused by device mirroring (extra 50\%, on average). The impact on memory consumption is minimal (extra 6\%, on average). Overall, memory does not appear to be an issue given less than 20\% utilization of the Raspberry Pi's 1~GB. The networking demand is also minimal, with just 32~MB of upload traffic for a $\sim$7 minutes test. Note that we set \texttt{scrcpy}'s video encoding (H.264) rate to 1~Mbps, which produces an upper bound of about 50~MB. The lower value depends on extra compression provided by \texttt{noVNC}. 

Evaluating the responsiveness of \tool's device mirroring is challenging. We call \emph{latency} the time between when an action is requested, either via automation or a click in the browser, and when the consequence of this action is displayed back in the browser, after being executed on the device. This depends on many factors like network latency (between browser and test device), load on device and/or controller, and software optimizations. We estimate such latency recording audio (44,100~Hz) and video (60~fps) while interacting with the device via the browser. We then manually annotated the video using ELAN multimedia annotator software~\cite{elan} and compute the latency as the time between a mouse click (identified via sound) and the first frame with a visual change in the app. We repeat this test 40 times while co-located with the vantage point (1~ms network latency) and measure an average latency of 1.44 ($\pm$0.12)~sec.

Next, we dig deeper into CPU utilization at the controller. Figure~\ref{fig:res-a} shows the CDF of the CPU utilization during the Chrome experiments with active and inactive device mirroring --- no significant difference was observed for the other browsers. When device mirroring is inactive, the controller is mostly underloaded, \ie constant CPU utilization at 25\%. This load is caused by the communication with the Monsoon to pull battery readings at highest frequency. When device mirroring is enabled, the median load instead increases to about 75\%. Further, in 10\% of the measurements the load is quite high and over 95\%.

\subsection{Location, Location, Location}
\label{sec:eval:vpn}
\begin{table}[t]
\small
\centering
\begin{tabular}{rcccc}
\hline
    {\bf Speedtest Server (kms)}   & {\bf D (Mbps)} & {\bf U (Mbps)} & {\bf L (ms)}\\
\hline
    \begin{tabular}{@{}c@{}}South Africa\\ Johannesburg (3.21)\end{tabular}  & 6.26  & 9.77  & 222.04 \\
    \hdashline[0.5pt/5pt]
    \begin{tabular}{@{}c@{}}China\\  Hong Kong (4.86)\end{tabular}           & 7.64  & 7.77  & 286.32 \\
    \hdashline[0.5pt/5pt]
    \begin{tabular}{@{}c@{}}Japan\\ Bunkyo (2.21)\end{tabular}               & 9.68  & 7.76  & 239.38 \\
    \hdashline[0.5pt/5pt]
    \begin{tabular}{@{}c@{}}Brazil\\ Sao Paulo (8.84)\end{tabular}           & 9.75  & 8.82  & 235.05 \\
    \hdashline[0.5pt/5pt]
    \begin{tabular}{@{}c@{}}CA, USA\\ Santa Clara (7.99)\end{tabular}        & 10.63 & 14.87 & 215.16\\
    \hdashline[0.5pt/5pt]
\end{tabular}
\vspace{0.1in}
\caption{ProtonVPN statistics. D=down/U=up/L=RTT.} 
\label{tab:vpn_summ}
\end{table}

\tool's distributed nature is both a \emph{feature} and a \emph{necessity}. It is a feature since it allows battery measurements under diverse set of network conditions which is, to the best of our knowledge, an unchartered research area. It is a necessity since it is the only way the platform can scale without incurring high costs. We here explore the impact of network location on battery measurements. In the lack of multiple \tool vantage points, we emulate such network footprint via a VPN. 

We acquired a basic subscription with ProtonVPN~\cite{protonvpn} and set it up at the controller. We then choose 5 locations from where to tunnel our tests. Table~\ref{tab:vpn_summ} summarizes the locations chosen, along with some statistics derived from SpeedTest (upload bandwidth, download bandwidth, and latency). VPN vantage points are sorted by download bandwidth, with the South Africa node being the slowest and the California node being the fastest. Since the speedtest server is always within 10~km from each VPN node, the latency here reported is mostly representative of the network path between the vantage point and the VPN node. 

\begin{figure}[tb]
   \centering
   \psfig{figure=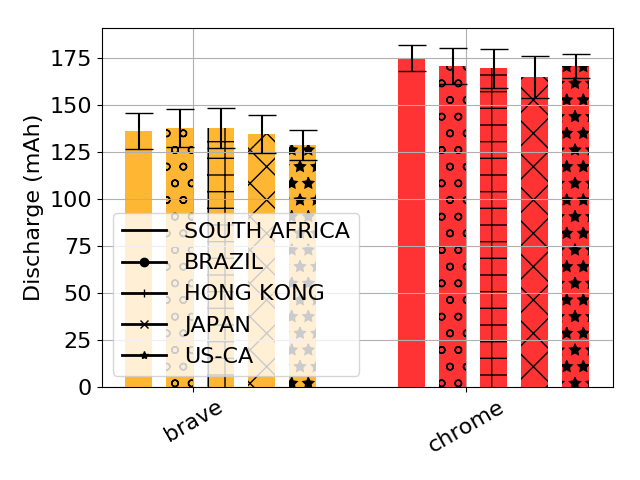, width=\columnwidth}
   \caption{Brave and Chrome energy consumption measured through VPN tunnels.}
   \label{fig:eval:vpn}
\end{figure}

Next, we extend the above automation script to also activate a specific VPN connection at the controller before testing. Figure~\ref{fig:eval:vpn} shows the average battery discharge (standard deviation as errorbars) per VPN location and browser --- for visibility reasons and to bound the experiment duration, only Chrome and Brave were tested. Overall, the figure does not show dramatic differences among the battery measurements as a function of the network location. For example, while the available bandwidth almost double between South Africa and California, the average discharge variation stays between standard deviation bounds. This is encouraging for experiments where \tool's distributed nature is a \emph{necessity} and its noise should be minimized. 

Figure~\ref{fig:eval:vpn} also shows an interesting trend when comparing Brave and Chrome when tested via the Japanese VPN node. In this case, Brave's energy consumption is in line with the other nodes, while Chrome's is minimized. This is due to a significant (20\%) drop in bandwidth usage by Chrome, due to a systematic reduction in the overall size of ads shown at this location. This is an interesting result for experiments where \tool's distributed nature is a \emph{feature}. 

Anecdotally, we also noticed that Google's lite pages\footnote{https://www.ghacks.net/2019/03/14/chrome-lite-pages/} were activated by default in South Africa and Japan, for Chrome. Google mentions that this decision is driven by low bandwidth rather than location, which does not necessarily match our measurements (see Table~\ref{tab:vpn_summ}). While we turned this feature off to ensure comparable tests, we also noticed that none of the tested pages currently support this feature. 

\section{Conclusion and Future Work}
\label{sec:conclusion}
In this paper we have proposed \tool, a distributed measurement platform for battery measurements. We have also started building and experimenting with \tool, to the point that our system is ready to accept new members. We specifically focused on Android because of ease of integration and availability of testing tools. However, we discussed iOS solutions which we soon plan to experiment with. Similarly, while we focus on mobile devices there is no fundamental constraint which would not allow \tool to support laptops or IoT devices. We designed \tool to enable remote access and human-controlled tests; we plan to facilitate such tests via integration with platforms like Mechanical Turk~\cite{mturk} and Figure Eight~\cite{figureeight}. Our vision is an open source and open access platform that users can join by sharing resources. However, we anticipate potential access via a credit system for experimenters lacking the resources for the initial setup.

\section*{Acknowledgments}
Katevas and Haddadi were partially supported by the EPSRC Databox and DADA grants (EP/N028260/1, EP/R03351X/1).

\bibliographystyle{abbrv} 
\begin{small}
\bibliography{biblio}
\end{small}
\end{document}